# Real-time Autonomous Robot for Object Tracking using Vision System

Qazwan Abdullah[1], Nor Shahida Mohd Shah[2], Mahathir Mohamad[3], Muaammar Hadi Kuzman Ali[4], Nabil Farah[5], Adeb Salh[6], Maged Aboali[7], Mahmod Abd Hakim Mohamad[8], Abdu Saif[9]

[1,4,6] Faculty of Electrical and Electronic Engineering, Universiti Tun Hussein Onn
[2] Faculty of Engineering Technology, Universiti Tun Hussein Onn Malaysia, Pagoh, Muar, Johor, Malaysia
[3] Faculty of Applied Science and Technology, Universiti Tun Hussein Onn Malaysia, Pagoh Education Hub, Pagoh, Muar, Johor
[5,7] Faculty of Electrical and Electronic Engineering Technology, Universiti Teknikal Malaysia Melaka
[8] Sustainable Product Development Research Group (SusPenD), Centre for Diploma Studies, Universiti Tun Hussein Onn Malaysia, Pagoh, Muar, Johor
[9] Faculty of Engineering, Universiti Malaya, Kuala Lumpur Malaysia

Email: [1]gazwan20062015@gmail.com, [3]mahathir@uthm.edu.my

*Abstract—* Researchers and robotic development groups have recently started paying special attention to autonomous mobile robot navigation in indoor environments using vision sensors. The required data is provided for robot navigation and object detection using a camera as a sensor. The aim of the project is to construct a mobile robot that has integrated vision system capability used by a webcam to locate, track and follow a moving object. To achieve this task, multiple image processing algorithms are implemented and processed in real-time. A mini-laptop was used for collecting the necessary data to be sent to a PIC microcontroller that turns the processes of data obtained to provide the robot's proper orientation. A vision system can be utilized in object recognition for robot control applications. The results demonstrate that the proposed mobile robot can be successfully operated through a webcam that detects the object and distinguishes a tennis ball based on its color and shape.

*Keywords--- Autonomous mobile robot, image processing, object tracking, mobile robot, vision-based navigation*

## I. INTRODUCTION

Mobile robots incorporating vision systems are the most desirable in their environment. They can move around freely, without being fixed to one particular physical position [1]. A mobile robot needs appropriate real-world information in real-time for improved navigation [2]. In addition, object tracking methods such as autonomous vehicle navigation, surveillance and many other applications are used in various robot vision applications. Although many studies and experiments have been carried out in recent decades, they have been concerned and committed to solving the problematic issues of monitoring the desired objective in chaotic and noisy environments [3]. The precision and versatility of intelligent robot solutions are improved by a robot with vision capability [4]. There exist many applications involving robots with vision systems, particularly in the manufacturing industry, research in science development, as well as space navigation purposes, because the vision system has provided numerous advantages [5]. A controller must be programmed to adjust the robot's movements in order to create an autonomous mobile robot equipped with a vision system (i.e.

camera) to navigate through the desired locations and locate desired objects [6]. By introducing this autonomous robot according to the images obtained, many advantages could be accomplished using image processing, such as securing buildings, maneuvering through dangerous environments, and so on. In this work, the Visual Basic programming language was used to program RoboRealm software [7]. Matlab was used to program the PIC microcontroller using a fuzzy logic concept [14-16]. In this paper, we developed a real-time autonomous robot for object tracking using the vision system by controlling a camera in the vision range of a robot that was used as a sensor in front of the robot, then by extracting camera information, it is used to actuate the motion of the robot.

## II. METHODOLOGY

Figure 1 depicts the system's overall layout. The device includes a camera attached to a mini-laptop as a vision sensor. The laptop is connected to the main controller through a motor driver circuit that interfaces with the robot motor using a PIC chip [8]. The mobile robot is programmed to find its path, monitor and use a webcam to follow an object, follow the path and monitor the object [9]. A control system helps the robot to keep tracking the object and follow it as it travels once the object is detected. The boards are mounted on the robot, including the camera that captures images which processes them through the compact microcontroller that adjusts the robot's motion and orientation according to the object's location. The robot must be capable to independently detect an object of interest and track it [10].

A web camera is used and mounted on the front of the robot. Due to cost constraints and for the purposes of small robots, it is intended to be as simple as possible [11]. A PC was used to be mounted on the top of the robot as the main controller for faster image processing, due to the computational complexity and extensive processing needed for image processing. This PC holds the decision-making algorithms. It performs the task of obtaining and processing image data from the camera, instructing the microcontroller to communicate with the positioning motors, and enabling the robot to move [12]. The PC is chosen because it provides a large competency of data processing results that can be visualized on the computer monitor, hence facilitating the fine-tuning process. To interpret, process and adjust the robot according to the incoming signals captured by the camera, a microcontroller is used. The microcontroller works as an interface to convert the computer system's serial data into a motor control signal to coordinate the robot's motors and drives. The robot's movements are determined by two stepper motors. The microcontroller controls them via a UCN5804B motor driver circuit to facilitate the programming section [13].

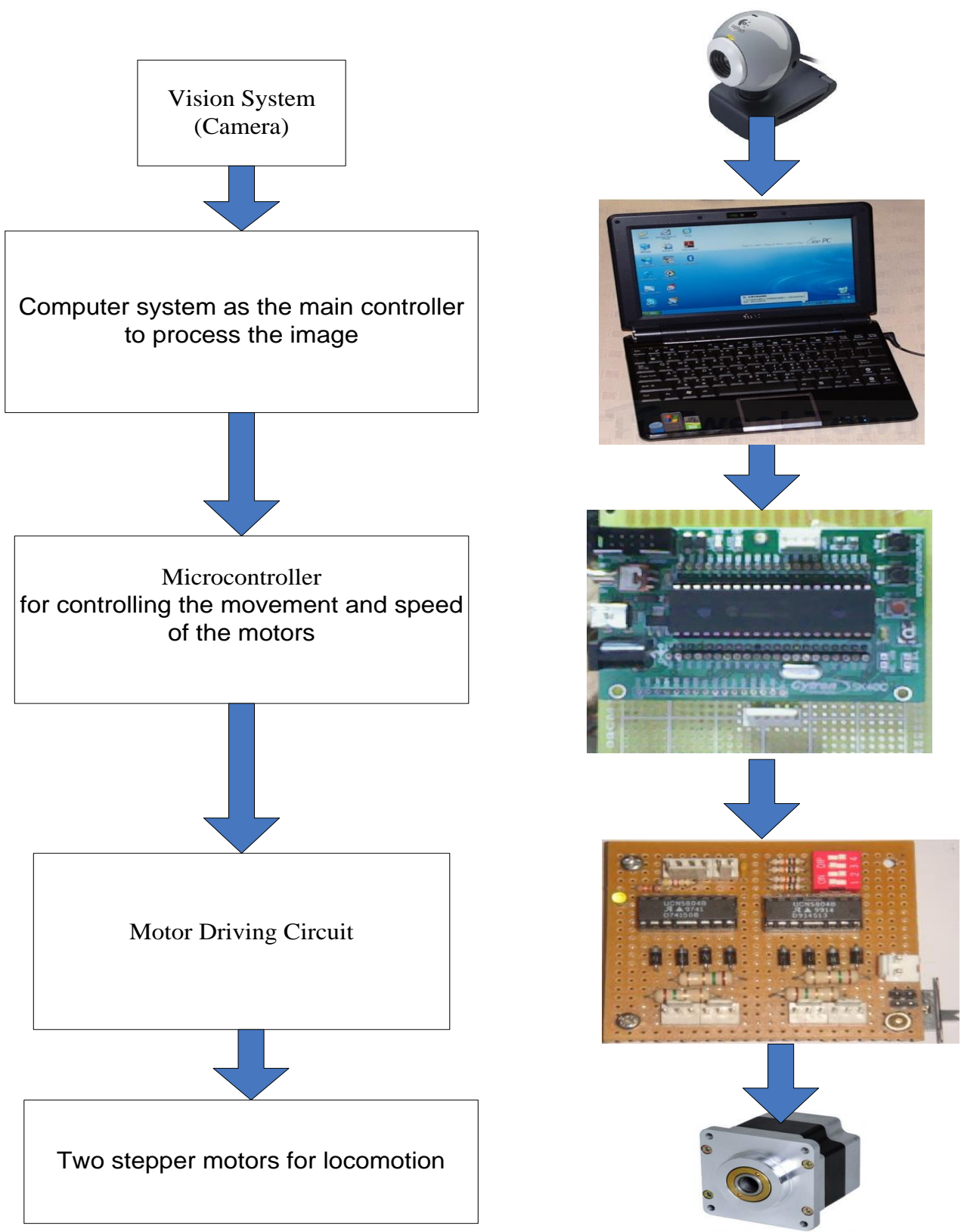

Figure 1. System Hardware Overview

To control the electronic instruments and drive motors, an appropriate power source supply was used. In order to enable the movement of the robot motors, the main circuit and circuit driver of the robot are supplied by 12V and ~2A battery.

**16280**

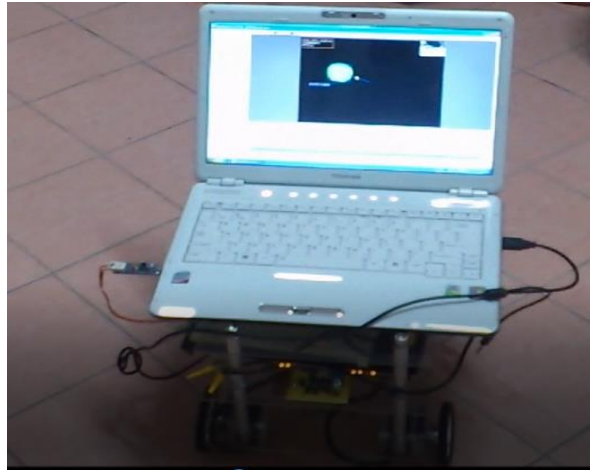

Figure 2. Robot front view with laptop.

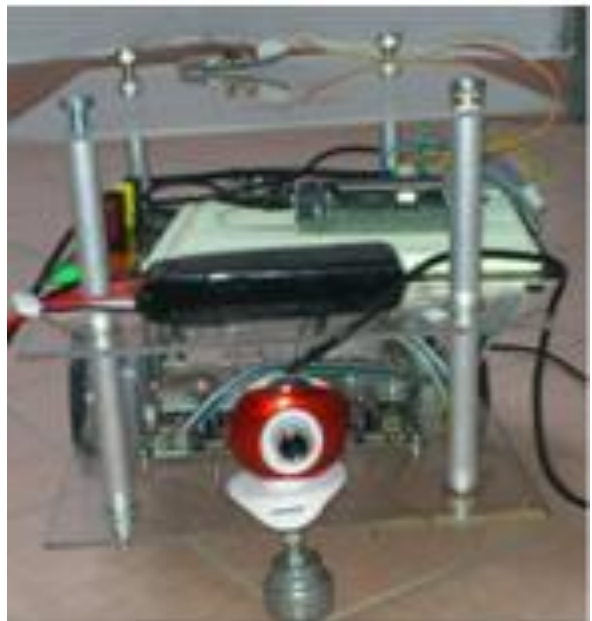

Figure3. Robot front view.

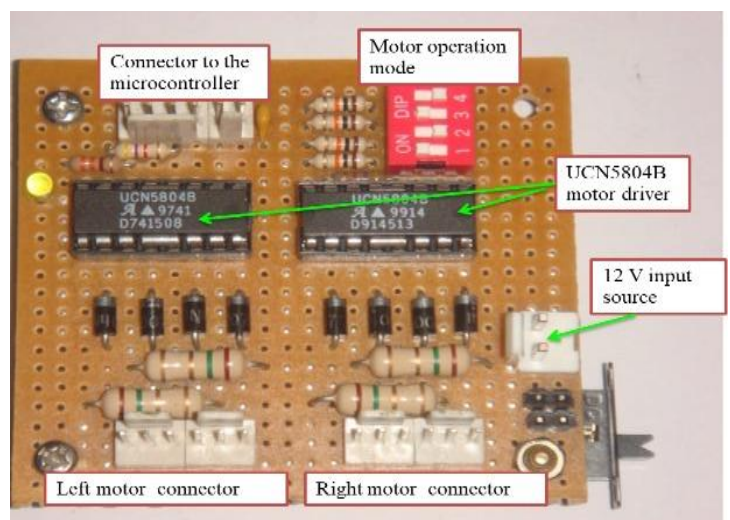

Figure 4. UCN5804B Motor Driver Actual Circuit

The three layers of the mobile robot are located 6 cm apart to allow the circuits to be mounted at a sufficient distance. The complete structure of the robot is shown in Figures 2 and 3. The main board is put on the top layer including the microcontroller and stepper motor controller circuit in the second layer. The stepper motors are also carried on both sides of the layer, and a mini laptop is put in the third layer. A cargo ball is mounted in the front part of the layer. The battery to power the robot is located in the second layer.

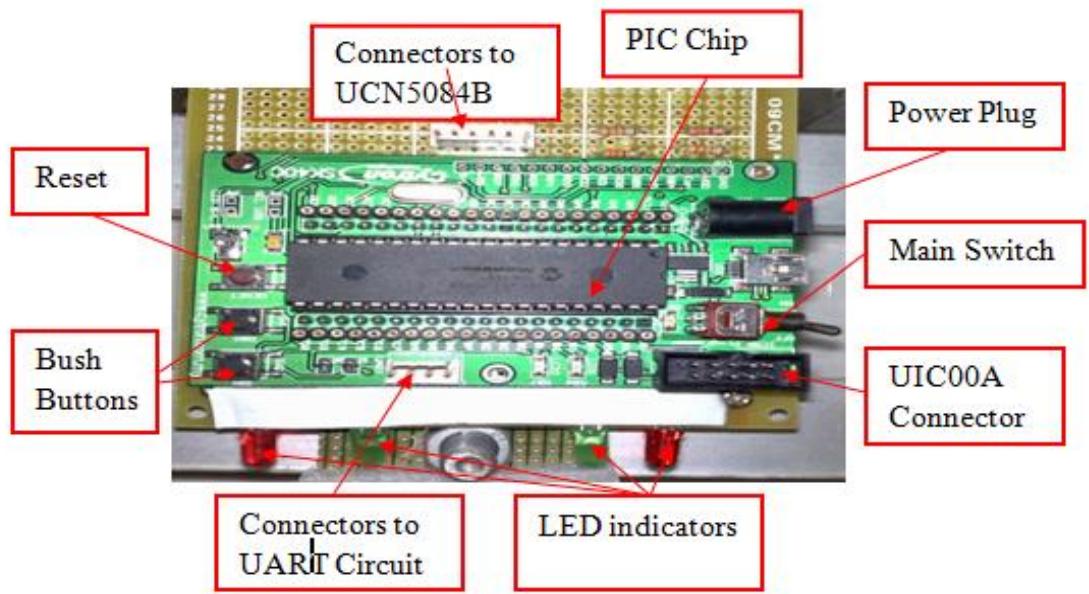

Figure 5. Main Board Circuit

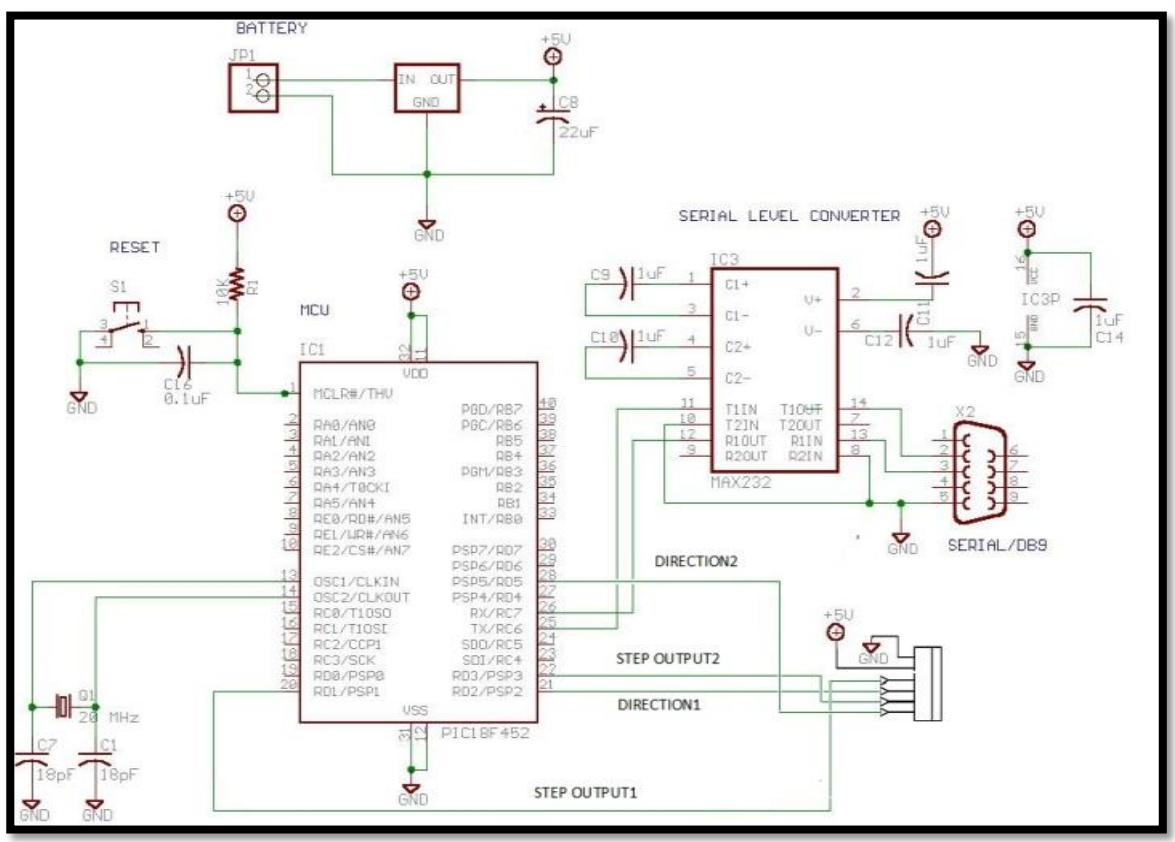

Figure 6. Main Board Schematic Circuit

The stepper motor controls the motor drivers shown in Figure 7, and consists of a UCN5804B controller for the stepper motor speed and direction control. There are three operating modes that can be selected according to the desired operation. For stepper motors, the two-phase operating mode is usually used, and the DIP switch is used to choose the appropriate mode of operation. Figure 7 shows the schematic view of the UCN5804B circuit, while Figure 4 shows the circuit 's actual configuration [17]. Two UCN5804B chips were used in both figures to independently power each motor. For the stepper motors, a 12 v power supply (battery) was used, while the circuit was operated by a 5 V source [7].

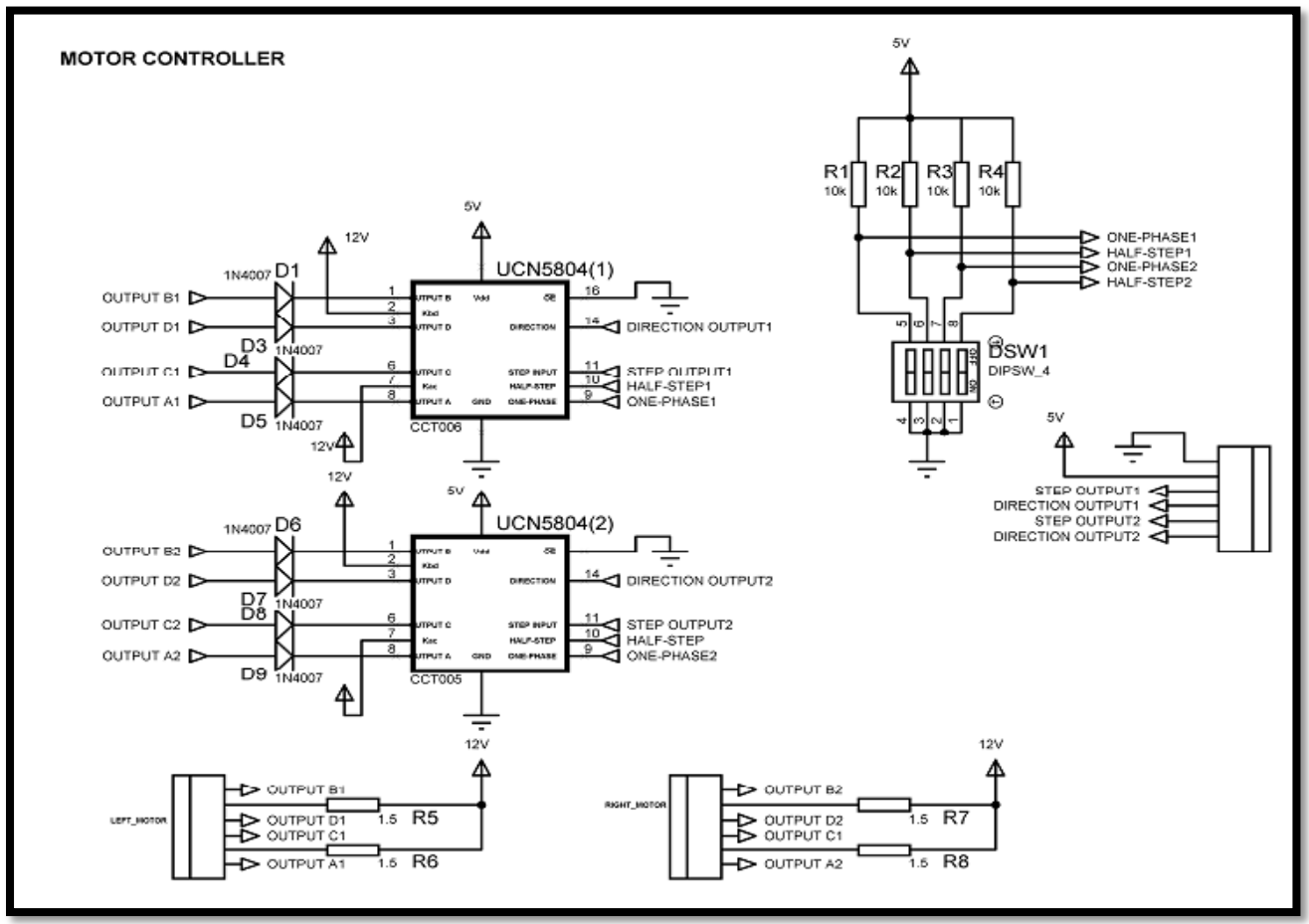

Figure 7. Motor Schematic Circuit

Image processing was used with the help of RoboRelm software with fast image real-time processing. It is resistant to different sizes, regardless of how far the object from the camera is. It uses a color filtering technique to consider the specific colors of objects, while other object colors are ignored. Since the object to be found is a tennis ball, the circular detection algorithm was used to check for the ball's centroid, even with some surface erosion of the circular form. If the ball is detected, an estimate of the ball's position is sent to the microcontroller, to give commands to the motor driver circuit to operate the motors to travel to the desired object and continue tracking it. Until the ball is detected, the robot is programmed to spin and search for the object if it disappears from the view of the camera. Figure 8 shows the image processing task.

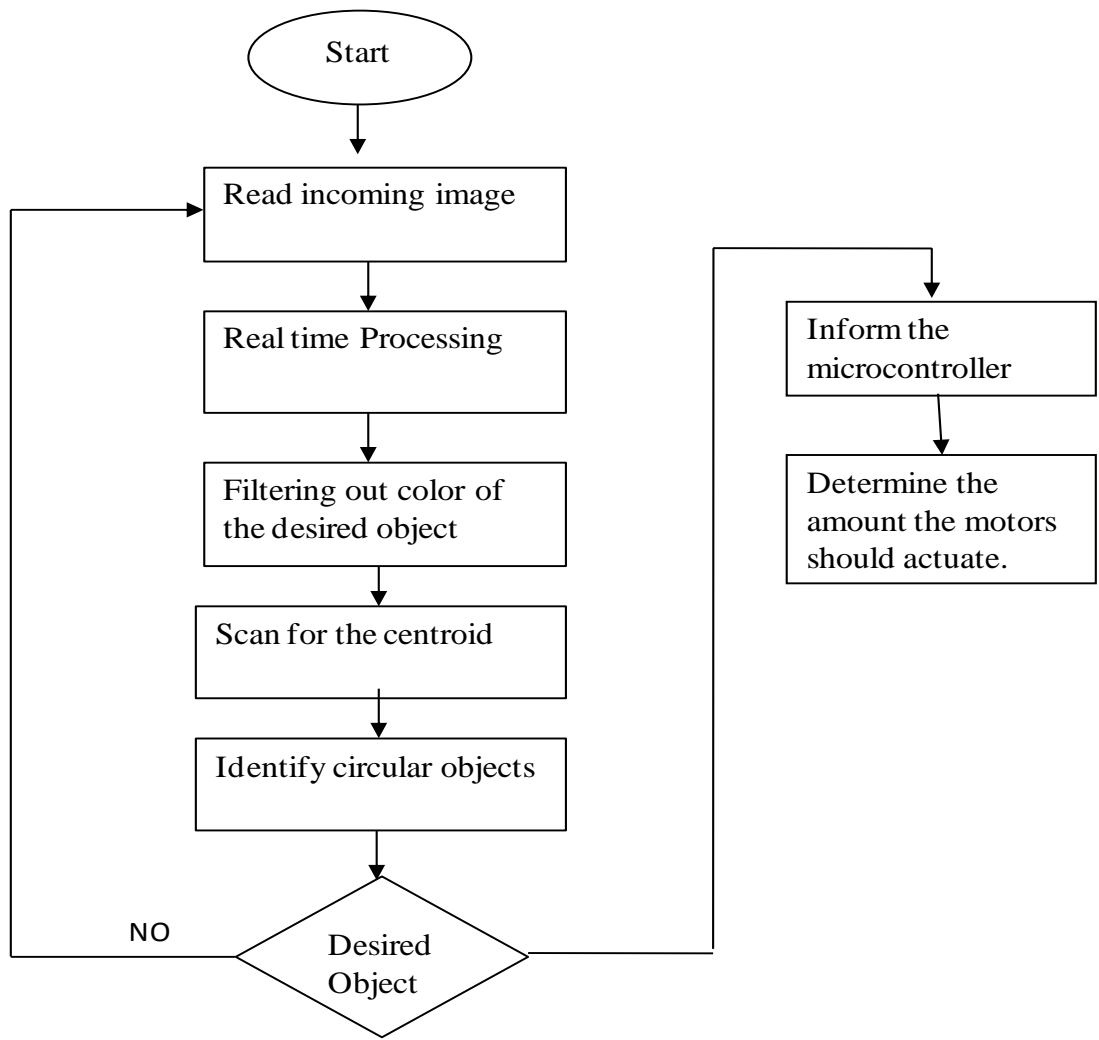

Figure 8. Image processing techniques flowchart

Figure 9 displays the PIC flowchart to regulate the direction and orientation of the motors. First, it specifies and initializes the input and output ports and the UART. Two red LEDs will blink to indicate that the initialization has been completed. It will wait for a moment to check if the data in the UART buffer is ready. The UART receives the data and initiates a selection loop to execute the appropriate command, based on the received data. Figure 9 shows six possible data items could be gathered with their corresponding actions. When the signal in the buffer is ready, the UART is read repeatedly [13].

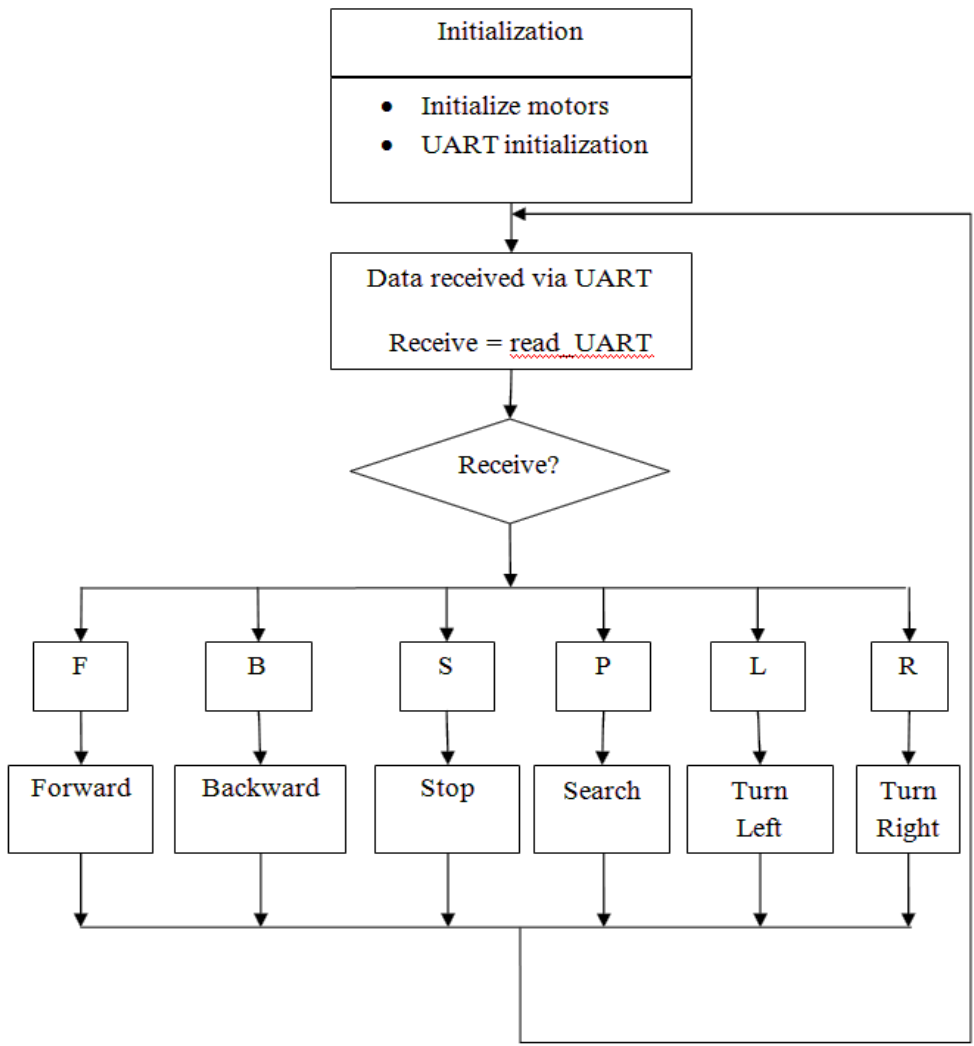

Figure 9. Control Decision Programming Flowchart

III. RESULTS AND ANALYSIS

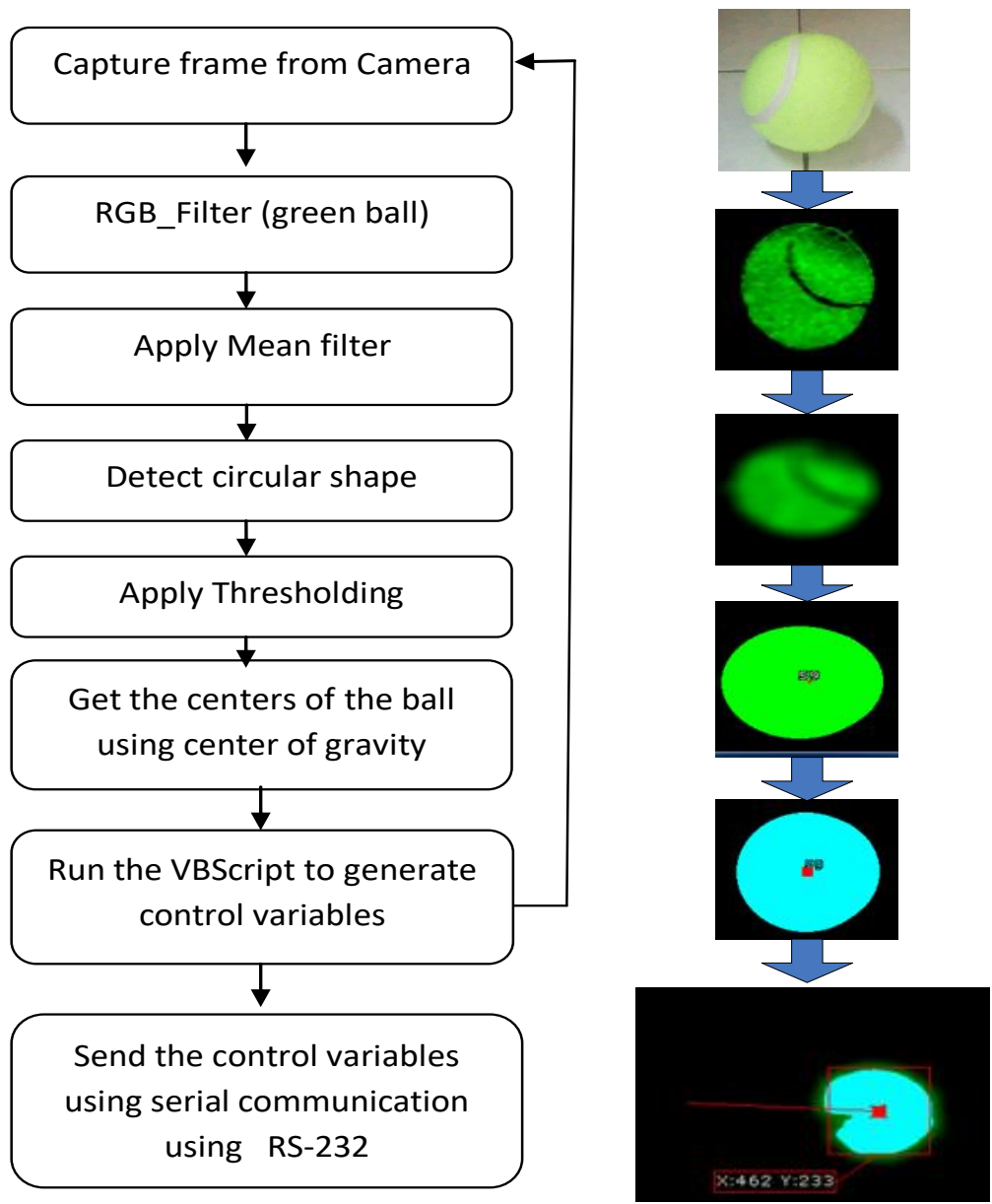

Figure 10. Image Processing Procedures

The complete ball detection process is as shown in Figure 10. The predefined custom color is green when only the green color channel is used, and any green color variation is caused by the other channels. The RGB filter in Roboleam's vision uses its values, and the filtration decreases all pixels that do not belong to the selected colors, depending on the particular color selected. This feature varies from the RGB channel in that white pixels are often diminished, even though the selected color may be included where

$$G = (G-R) + (G-B) \text{ and } R=0, B=0 \qquad (1)$$

The pixels with dark or white colors are omitted using a thresholding property that does not contain a sufficient green color. Any green noises caused by changes in lighting or intensity are removed by a degree of hysteresis that is applied to the object. Figure 12 shows the segmentation of the green ball from the background for further processing. The noises created by additional factors such as lighting intensity and the presence of moving particles are filtered out by a mean or average filter. By taking the average of pixels, the object is further smoothed and blurred, which has the benefit of removing any unwanted differences in noise factors. Figure 13 shows the resulting object. It can be shown that the green pixels in Figure 12 are connected by comparing Figures 11 and 12, removing any gaps between neighboring pixels or any other pixels that are not part of the ball. It also helps to eliminate textures beyond the ball's diameter [13].

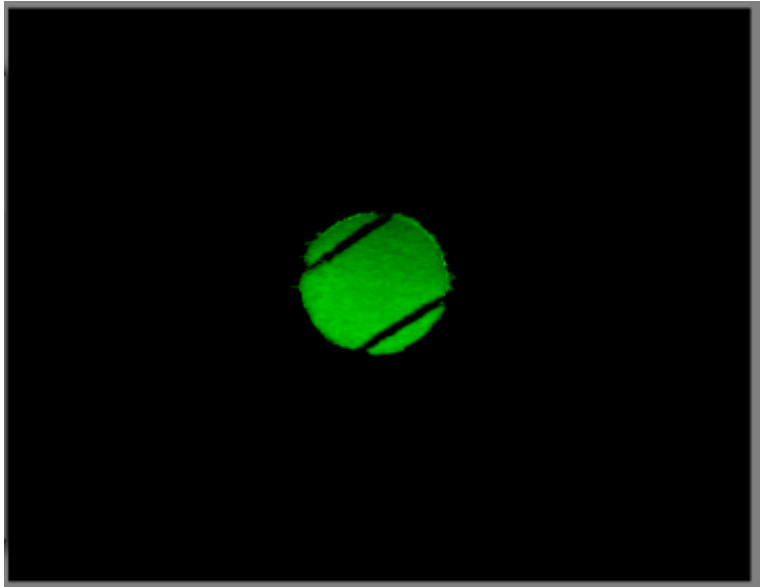

Figure 11. Ball Color Detection

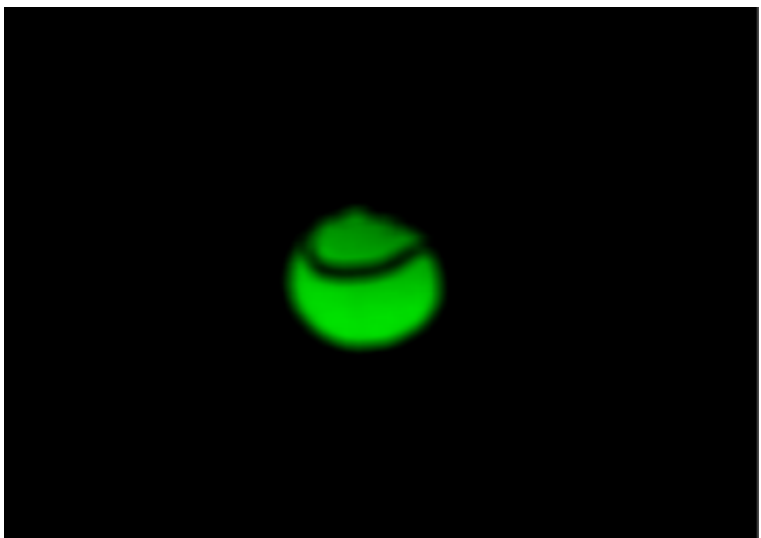

Figure 12. Noise Reduction

The shape of the ball represents the image using circular shape properties to lead to an absolute spherical ball. The circular shape is detected based on the selected threshold angle. In addition, the radius is specified within a specific range of 10-210 pixel values. The pixels that are not part of the object are ignored by applying point clustering and linking pixels that are similar to each other to group objects. It is a method of discarding any pixels that are not part of the ball by adding a limit. The product of point clustering is shown in Figure 13.

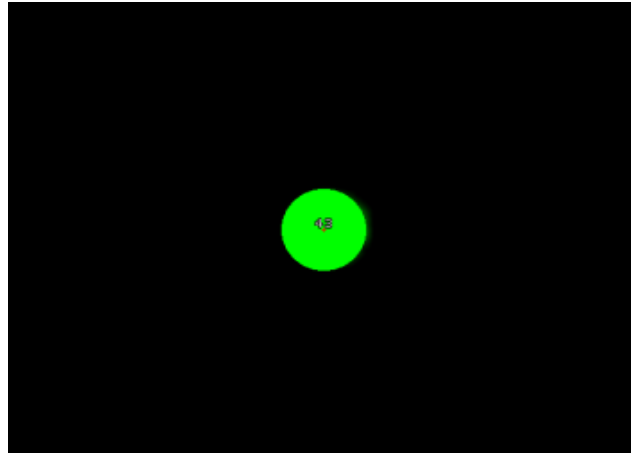
Figure 13. Circle Detection

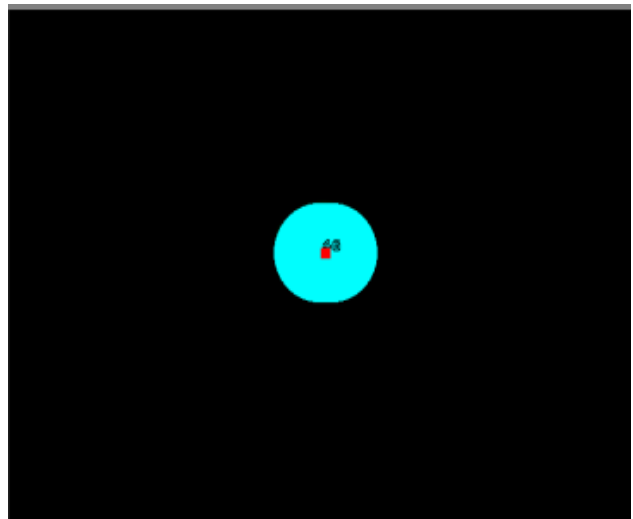
Figure 14. Points Clustering

Figure 14 shows the eliminated portions of the image after a thresholding process has been passed to eliminate portions of the image that fall within a defined range of intensity. The image is divided into parts that have consistent intensity values. Only green color pixels are selected, and the other colors are deleted. Lighting conditions are considered to avoid the need to adjust the parameters as the lighting variables change. The robot may also be used in various environments with varying lightening conditions. The center of gravity is used to locate the coordinates of the ball's center, which allows the robot to align it itself according to its coordinates, which are considered the most sensual part image processing. The ball centroid coordinates are shown in X and Y, as shown in Figure 16. The location of the ball is required to be identified by using the X coordinate, and whether the ball is in the left or right of the reference point. This is calculated as follows:

$$X = \frac{\sum_{k=0}^{n} x_n}{n} \qquad (2)$$

where x_n is the coordinate of the n$^{th}$ pixel, and n is the total pixels in the chosen object. The box encircles the ball, and its size is important to evaluate how far the ball is from the camera. The robot will go backwards and stop if the size of the box is large to avoid hitting the ball.

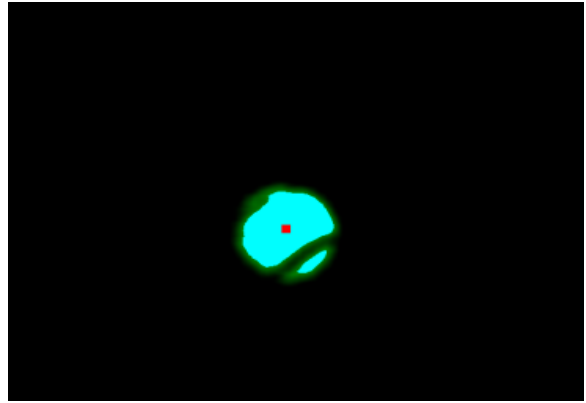

Figure 15. Thresholding Process.

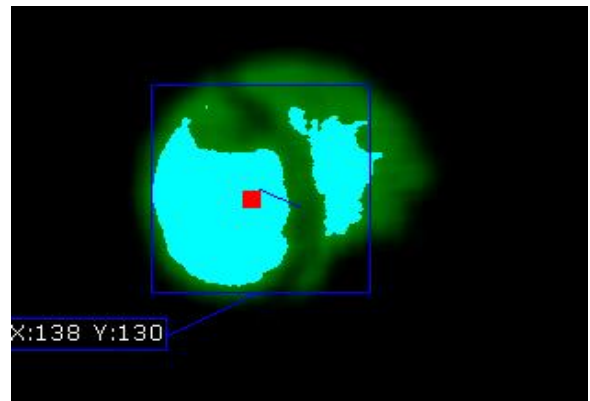

Figure 16. Center of Gravity.

The mobile robot's orientation and direction are centered on the ball's location, relative to the reference point indicated by the image's mid-point (640 by 480). Figures 17 to 22 display the six different potential mobile robot directions. If the robot camera does not locate the ball, the search mode is set by spinning on the spot before the ball is located, as shown in Figure 18. When the RoboRealm program gives the order "P" to the microcontroller, the robot starts the "navigation" mode. Otherwise, the robot moves forward to the ball within a certain value range of 70 pixels to the left or right of the center of the frame if the RoboRealm gives the microcontroller order of 'F', as shown in Figures 17 and 18.

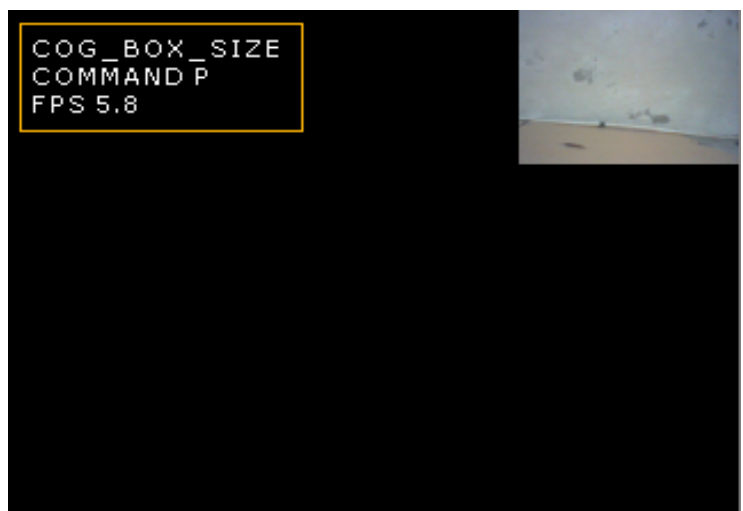

Figure 17. Searching for Ball

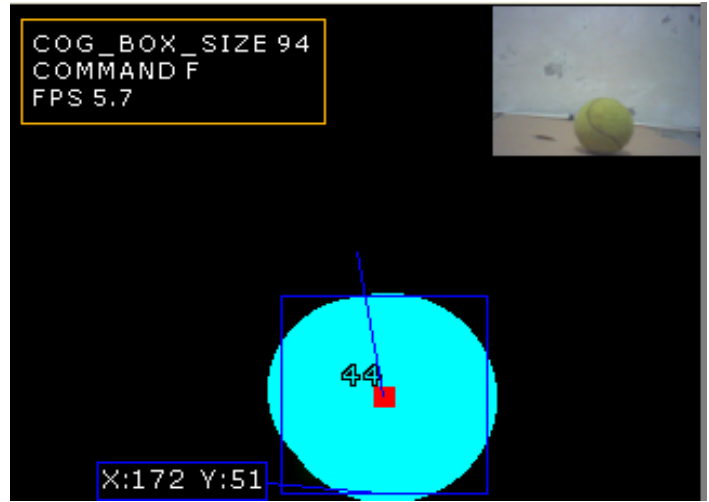
Figure18. Forward

The box size determines whether the ball is close to the robot. If the robot is sufficiently in close proximity to the ball, the PIC receives an "S" signal to inform the robot to stop moving. The robot will stop due to the location of ball if the size of the box is 136 with the range 130 to 130 + 100. The robot moves backwards and prevents the ball from crashing if the ball approaches the robot within a range greater than 230. As a consequence, the PIC receives the "B" signal, as shown in the Figure 19.

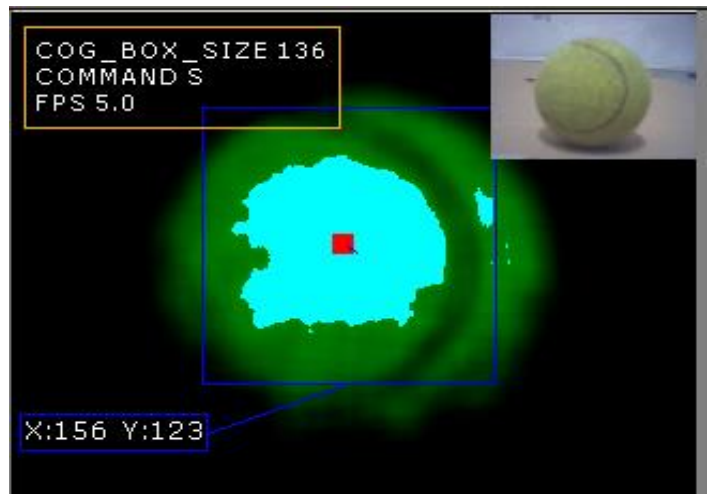
Figure 19. Stop Position

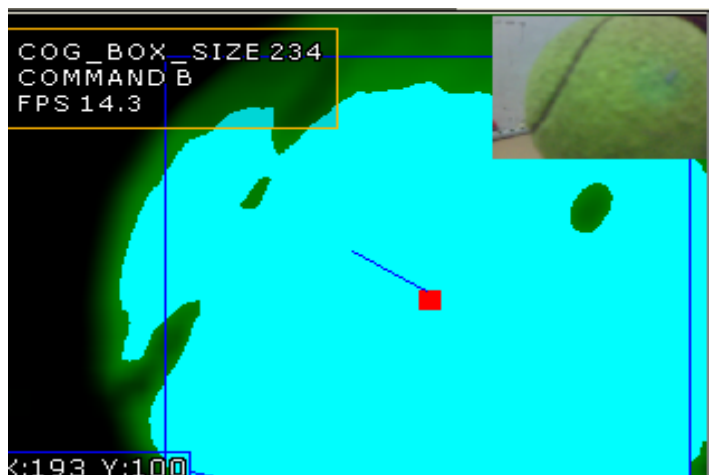
Figure 20. Backward Position

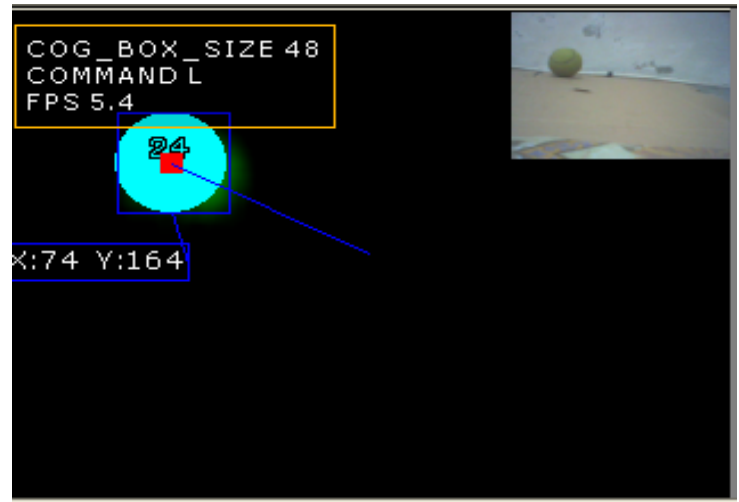

Figure 21. Left Direction

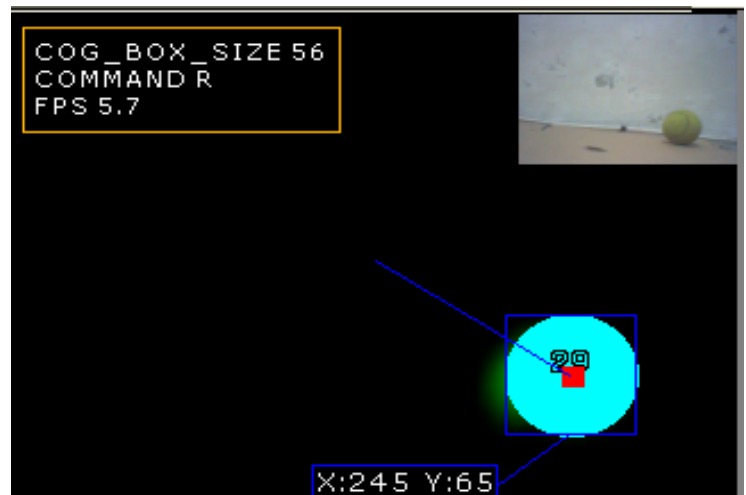

Figure 22. Right Direction

The PIC receives the command "L" if the ball is on the left, and command "R" if it is on the right, as shown in Figures 22 and 23 respectively. The PIC gives the order to the motors to travel left and right, tracking the ball. The box size surrounds the ball within the range of 10 to about 130, to make the robot move forward, left or right, and also makes the robot stop or move backwards. The size of the box is needed for calculation of movement.

## IV. CONCLUSION

The vision system object tracking robot was successfully accomplished using only a webcam as the main object detection sensor, demonstrating a great ability to distinguish a tennis ball based on the color and shape, and track the ball as it travels in any direction. The robot has been fitted with the mechanism to search for the ball, and keeps monitoring it if the ball is not present in the view of the camera by spinning in place until the ball is detected. Extensive image processing techniques and algorithms need to be processed on-board using a mini-laptop for rapid processing to accomplish the task. The interpreted information is transmitted to the microcontroller, and converted into real world orientation.

V. ACKNOWLEDGMENTS

The authors would like to thank the Ministry of Higher Education Malaysia and Universiti Tun Hussein Onn Malaysia for supporting this research under the Fundamental Research Grant Scheme (FRGS) vot FRGS/1/2018/TK04/UTHM/02/14 and TIER1 Grant vot H158.

## AUTHORS PROFILE

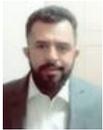
**Qazwan Abdullah** was born in Taiz, Yemen and received his bachelor's degree of Electrical and electronic engineering from Universiti Tan Hussein Onn Malaysia (UTHM) in 2013. Also received his Master of Science in Electrical and Electronic Engineering in 2015 from the same university. Currently, he is a PhD student at the Faculty of Electrical & Electronic Engineering with research interests of control system, wireless technology and microwave

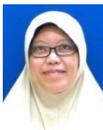
**Nor Shahida Mohd Shah** received B.Eng, in Electrical and Electronic from Tokyo Institute of Technology, Japan in 2000. She then received M.Sc. in Optical Communication from University of Malaya, Malaysia in 2003. She received PhD in Electrical, Electronics and System Engineering from Osaka University, Japan in 2012. She is currently a Senior Lecturer at Universiti Tun Hussein Onn Malaysia. Her research interests are optical and wireless communication.

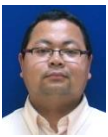
**Mahathir Mohamad** received the B.Sc. degree from Universiti Putra Malaysia, in 2000, the M.Sc. degree from the National University of Malaysia, in 2003, and the Ph.D. degree from Osaka University, in 2012. He currently holds the position as a Senior Lecturer at Universiti Tun Hussein Onn Malaysia. He is majoring in applied mathematics.

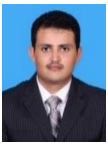
**Muaammar Hadi Kuzman Ali** was born in Yemen in 1988. He has received a bachelor's degree in electrical engineering (robotic) from Universiti Teknologi Malaysia in 2011. And received a master's degree in electrical engineering at the same university. Research interest is the robotic

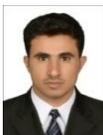
**N. Farah** was born in Yemen in 1988. He has received a bachelor's degree in electrical engineering (Power electronics and drives) from Universiti Teknikal Malaysia Melaka in 2015.And received master's degree in electrical engineering in same university. Currently he is enrolled as PhD student in electrical engineering also at same university. Current research interest is self-tuning fuzzy logic controller of AC Motor drives and predictive control of induction motor drives.

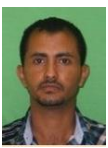
**A. Salh** currently a postdoctoral researcher at Electrical and Electronic Engineering, Universiti Tun Hussein Onn Malaysia, Received the Bachelor of Electrical and Electronic Engineering, IBB university, IBB, Yemen (2007), and the Master and PhD of Electrical and Electronic Engineering, University Tun Hussein Onn Malaysia, Malaysia (201**5**& 2020).

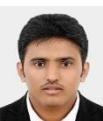
**MAGED ABOALI** was born in Yemen, in 1993. He received the B.Eng. degree (mechatronics with Hons.) from the Universiti Teknikal Malaysia Melaka, in 2015, and the M.S. degree (Hons.) in electronic engineering from the Universiti Teknikal Malaysia Melaka, in 2018. His main research interests include artificial intelligent, image processing, computer vision, speech synthesis, stereo vision, AC motor drives and control of induction motor drives.

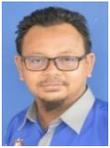

**Mahmod Abd Hakim Mohamad** is a lecturer at the Department of Mechanical Engineering, Centre for Diploma Studies, University of Tun Hussein Onn Malaysia (UTHM), since 2011. In June 1996, he graduated with Certificate in Mechanical Engineering. After that, he continued his studies and after one year, he got his diploma in the same field from Port Dickson Polytechnic. He furthered his tertiary education in Kolej Universiti Teknologi Tun Hussein Onn (KUiTTHO) and obtained a Bachelor's Degree in Mechanical Engineering with honors in 2002 and after 2010 complete study at University of Putra Malaysia (M. Sc.) in Aerospace Engineering and started to get involved in research field of computational fluid dynamics. He is active in research and innovation products in the field of teaching and learning, especially in design engineering. He also actively published various journals and conference papers

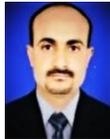

**Abdu Saif** received his Master of Science degree in Project Management in 2017 from University of Malaya, Malaysia and B.E. Degree in Electronics communication in 2005 from IBB University, Yemen. He has more than 9 years of industrial experience in telecommunications companies. He is currently pursuing his Doctorate Degree in Electrical Engineering (major in wireless communication) from the Faculty of Engineering, University of Malaya, Kuala Lumpur, Malaysia. His research interests include Wireless networks, 3D coverage by UAV, Internet of flying things, emergency management system, and public safety communication for 5G.